\documentclass[a4paper,fleqn]{cas-dc}

\usepackage[numbers]{natbib}
\usepackage{hyperref}
\hypersetup{
	colorlinks=true,
	linkcolor=cyan,
	filecolor=blue,      
	urlcolor=red,
	citecolor=blue,
}

\def\tsc#1{\csdef{#1}{\textsc{\lowercase{#1}}\xspace}}
\tsc{WGM}
\tsc{QE}
\tsc{EP}
\tsc{PMS}
\tsc{BEC}
\tsc{DE}

\begin{document}
\let\printorcid\relax
\let\WriteBookmarks\relax
\def\floatpagepagefraction{1}
\def\textpagefraction{.001}

\shorttitle{}

\shortauthors{Lou et al.}

\title [mode = title]{SDR-Former: A Siamese Dual-Resolution Transformer for Liver Lesion Classification Using 3D Multi-Phase Imaging}                      

\author[1,4]{Meng Lou}
\fnmark[1]
\author[2]{Hanning Ying}
\fnmark[1]
\author[1]{Xiaoqing Liu}
\fnmark[1]
\author[4]{Hong-Yu Zhou}
\author[3]{Yuqing Zhang}
\cormark[1]
\author[4]{Yizhou Yu}
\cormark[2]

\address[1]{Artificial Intelligence Laboratory, Deepwise Healthcare, Beijing, China}
\address[2]{Department of General Surgery, Sir Run Run Shaw Hospital, Zhejiang University School of Medicine, Hangzhou, Zhejiang, China}
\address[3]{Department of Radiology, Ningbo Medical Center Lihuili Hospital, Ningbo, Zhejiang, China}
\address[4]{Department of Computer Science, The University of Hong Kong, Pokfulam, Hong Kong}

\fntext[fn1]{Authors contribute equally to the paper.}
\cortext[cor1]{Corresponding authors: Yizhou Yu (yizhouy@acm.org) and Yuqing Zhang (zyqlucky168@163.com)}
\cortext[cor2]{Corresponding author at: Department of Computer Science, The University of Hong Kong, Pokfulam, Hong Kong}

\begin{abstract}
Automated classification of liver lesions in multi-phase CT and MR scans is of clinical significance but challenging. This study proposes a novel Siamese Dual-Resolution Transformer (SDR-Former) framework, specifically designed for liver lesion classification in 3D multi-phase CT and MR imaging with varying phase counts. The proposed SDR-Former utilizes a streamlined Siamese Neural Network (SNN) to process multi-phase imaging inputs, possessing robust feature representations while maintaining computational efficiency. The weight-sharing feature of the SNN is further enriched by a hybrid Dual-Resolution Transformer (DR-Former), comprising a 3D Convolutional Neural Network (CNN) and a tailored 3D Transformer for processing high- and low-resolution images, respectively. This hybrid sub-architecture excels in capturing detailed local features and understanding global contextual information, thereby, boosting the SNN's feature extraction capabilities. Additionally, a novel Adaptive Phase Selection Module (APSM) is introduced, promoting phase-specific intercommunication and dynamically adjusting each phase's influence on the diagnostic outcome. The proposed SDR-Former framework has been validated through comprehensive experiments on two clinical datasets: a three-phase CT dataset and an eight-phase MR dataset. The experimental results affirm the efficacy of the proposed framework. To support the scientific community, we are releasing our extensive multi-phase MR dataset for liver lesion analysis to the public. This pioneering dataset, being the first publicly available multi-phase MR dataset in this field, also underpins the MICCAI LLD-MMRI Challenge. The dataset is accessible at: \textcolor{blue}{https://bit.ly/3IyYlgN}.
\end{abstract}




\maketitle

\section{Introduction}
Combating liver cancer remains a significant challenge in global health, demanding immediate progress in both diagnosis and treatment methods\cite{ferlay2021cancer}. Presently, contrast-enhanced multi-phase computed tomography (CT) and magnetic resonance (MR) imaging are the primary techniques for diagnosing liver lesions \cite{lee2020ct}. These imaging methods provide extensive anatomical and functional insights, from vascular patterns to lesion features, enhancing their accuracy and reliability in lesion identification \cite{quatrehomme2012assessing}. Nevertheless, analyzing 3D multi-phase images manually is a demanding and challenging task, even for experienced radiologists. This difficulty is further exacerbated by the variability in lesion appearances, indistinct lesion margins, and the unique characteristics presented in each imaging phase, all of which make lesion classification more complex \cite{zheng2018unified, luo2022rare}. These challenges highlight the urgent need to develop a precise and resilient computer-aided diagnostic (CAD) system, aiming at assisting radiologists in providing more accurate and rapid diagnostic assessments.
\par
Deep learning has rapidly risen to prominence in medical image analysis, primarily due to its exceptional capacity to encapsulate intricate representations of medical images \cite{zhou2021review}. This advancement is particularly evident in the domain of multi-phase medical image analysis, where state-of-the-art deep learning techniques have been instrumental in achieving groundbreaking results \cite{yasaka2018deep,khan2020multimodal,raju2020co,wang2021transbts,liang2018combining,jiang2020multi,qu2022m3net}. Presently, the field of multi-phase medical image analysis is predominantly guided by two influential approaches: image-level fusion and feature-level fusion. The image-level fusion approach is notable for its scalability and computational efficiency. It treats each phase as a separate channel in a unified input image \cite{wang2021transbts}, allowing for seamless adaptation to datasets with different numbers of phases while conserving computational resources. Additionally, this approach paves the way for transfer learning across datasets with different phase counts, facilitating the application of insights gained from one dataset to enhance outcomes on another. Conversely, by setting up a dedicated model for each phase, the feature-level fusion approach is distinguished by its ability to thoroughly harness multi-phase information, promoting detailed interaction among features from various phases \cite{liang2018combining,jiang2020multi,qu2022m3net}. This approach enhances the model's capacity to represent multi-phase characteristics, resulting in improved performance.
\par
However, both approaches come with intrinsic limitations. The image-level fusion approach has a risk of diluting the richness of feature representation within individual phases. This often stems from the preliminary fusion of multi-phase images at early stages, potentially limiting the model's representational capabilities and leading to sub-optimal performance. In contrast, feature-level fusion faces distinct hurdles. Notably, assigning a separate model for each phase results in considerable computational demands, which may surpass hardware capacities as the number of phases increases. This high computational load can hinder the practicality of this approach, both in model training and in actual clinical settings. Additionally, the adaptability of the model is hindered when dealing with datasets having different phase counts. This limitation affects the smooth implementation of transfer learning and necessitates extra modifications to ensure model compatibility with varying phase counts.
\par
To overcome the inherent drawbacks of both image-level and feature-level fusion approaches, we propose the Siamese Dual-Resolution Transformer (SDR-Former), a novel framework designed for liver lesion classification in 3D multi-phase CT and MR imaging with varying phase counts. Our framework utilizes a Siamese Neural Network (SNN) for handling multi-phase inputs, offering two major benefits: (1) By allocating an individual neural network to each phase, our method surpasses the image-level fusion approach in feature representation capabilities; (2) Through the use of a weight-sharing network for encoding multi-phase images, our proposed framework achieves exceptional scalability and computational efficiency, effectively addressing the limitations of the feature-level fusion approach. However, our experiments revealed limitations when employing a CNN-based SNN, attributed to the static weight generation of CNN models, limiting their representation capabilities. To enhance the feature representation power of the SNN, we design a hybrid CNN-Transformer network named the Dual-Resolution Transformer (DR-Former), which combines the strengths of CNNs and Transformers: it utilizes a 3D CNN for detailed local feature extraction from high-resolution images and an efficient 3D Transformer for global context analysis from low-resolution images. We also introduce a Bilateral Cross-resolution Integration Module (BCIM) to enable semantic interactions and extract complementary features from the dual-resolution feature maps produced by the DR-Former. Additionally, we propose an Adaptive Phase Selection Module (APSM) to enhance inter-phase contextual communication and dynamically adjust the influence of each phase in the final diagnostic classification. 
\par
In summary, our contributions are multifaceted:
\begin{itemize}
    \item We introduce the SDR-Former, a versatile framework tailored for liver lesion classification in 3D multi-phase CT/MR imaging. This framework stands out for its effective multi-phase feature extraction, exceptional scalability, and computational efficiency, effectively overcoming the constraints of traditional fusion techniques.
    \item To augment the feature representation capabilities of the SNN, we design a unique hybrid CNN-Transformer network, namely DR-Former. Additionally, we introduce the APSM, designed to dynamically adjust the relevance of each imaging phase, thereby enhancing diagnostic accuracy and contextual sensitivity.
    \item We thoroughly validate our SDR-Former framework using two clinical datasets: a three-phase CT dataset with two lesion types and an eight-phase MR dataset with seven different lesion categories. Our experimental results confirm the superiority of the SDR-Former, establishing it as a cutting-edge solution in liver lesion classification.
    \item As a contribution to the research community, we are making our comprehensive 3D multi-phase MR dataset available publicly. The complete dataset will be accessible post-manuscript approval. To the best of our knowledge, this is the first public MR dataset that encompasses up to eight imaging phases with seven lesion categories, dedicated to liver lesion diagnosis.
\end{itemize}
\par
The rest of the paper is organized as follows: Section~\ref{sec:related_work} reviews relevant prior studies. Section~\ref{method} details the proposed methodology. Section~\ref{experiment_set} outlines the experimental design, and Section~\ref{results} analyzes the experimental results. The paper concludes in Section~\ref{conclusion}, summarizing our primary findings and contributions.

\section{Related Work}
\label{sec:related_work}
\subsection{Information Exploitation in Multi-phase Imaging}
The abundant information contained in multi-phase images has led to significant strides in their analysis, thanks to the application of diverse deep learning techniques. Liang et al. \cite{liang2018combining} adapted ResNet into both global and local processing pathways, incorporating a bidirectional long short-term memory (LSTM) for efficient classification of focal liver lesions in multi-phase CT images. Jiang et al. \cite{jiang2020multi} developed an advanced network utilizing multi-phase and multi-level feature integration, enabled by dual feature encoders and an elaborate adaptive cross-refinement module, thereby enhancing pancreas segmentation accuracy in CT imaging. Wang et al. \cite{wang2021transbts} successfully integrated a Transformer with a 3D CNN for multi-modal MR image analysis, achieving significant improvements in brain tumor segmentation by fusing multi-modal data at the image level. Recently, Qu et al. \cite{qu2022m3net} introduced M3Net, a multi-scale, multi-view approach for CT-based multi-phase pancreas segmentation. The authors employed a non-local attention mechanism \cite{wang2018non} to improve inter-phase feature connections, resulting in effective segmentation of the pancreas in varied pathological conditions. 

\subsection{Incorporating Multi-resolution Data in Image Analysis}
Leveraging multi-resolution inputs is pivotal for capturing both coarse- and fine-grained information, a principle that has led to various advancements in medical image analysis. Xu et al. developed a multi-scale, cost-sensitive neural network (MSCSNN), employing three parallel ResNets to process images at different resolutions, thereby generating multi-scale representations for accurately detecting malignant lung nodules \cite{xu2020mscs}. HookNet utilizes dual CNNs to separately process low- and high-resolution images, combining detailed and broader features for semantic segmentation in histopathology images \cite{van2021hooknet}. In the semi-supervised reverse adversarial network by Fu et al., a multi-resolution CNN is used to enhance the interaction between inputs of varying resolutions, improving the diagnostic precision of pulmonary nodules \cite{fu2022semi}. In this work, our hybrid CNN-Transformer network (i.e., DR-Former) capitalizes on this multi-resolution concept. It utilizes a 3D CNN for analyzing high-resolution images and a 3D Transformer to handle low-resolution images, based on the idea that convolutional and self-attention mechanisms serve as respective high- and low-pass filters and can capture unique spatial details crucial \cite{park2022vision}.

\subsection{Siamese Neural Networks}
Siamese Neural Networks (SNNs), a distinct category of neural network designs, are tailored for analyzing paired inputs through a shared-weights mechanism \cite{bromley1993signature}. Their adaptability has led to diverse applications in fields like object tracking \cite{fu2021stmtrack}, few-shot image classification \cite{li2023deep}, and the development of invariant visual representations using contrastive learning \cite{chen2020simple,chen2021exploring,wang2022few}. In the realm of multi-phase medical imaging, we propose that SNNs hold considerable promise for the effective extraction of features from multi-phase imagery, a prospect that our study seeks to investigate for liver lesion classification in multi-phase CT and MR scans. By utilizing an SNN framework, we endeavor to not only extract robust features from each imaging phase but also develop a model that is flexible and scalable across datasets with varying phase counts.

\subsection{Advancements in Vision Transformer Models}
Drawing inspiration from Transformer models in natural language processing (NLP) \cite{vaswani2017attention}, Vision Transformer (ViT) models have ushered in a new era in computer vision. Similar to NLP tokens, ViTs dissect images into patch sequences and use self-attention mechanisms to understand the global relationships between these patches \cite{dosovitskiy2021image}. Although traditional ViTs have achieved impressive results, they tend to be resource-intensive, particularly with high-resolution images. To address these computational demands, various refined Transformer models have been developed. The Swin Transformer introduces a method of creating hierarchical image structures and implementing self-attention in shifting local windows across layers \cite{liu2021swin}. Pyramid Vision Transformer (PVT) manages computational efficiency by reducing the spatial dimensions of key-value pairs at different stages \cite{wang2021pyramid}. Pyramid Pooling Transformer (P2T) streamlines self-attention by condensing key-value data through multi-scale pooling \cite{wu2021p2t}. Furthermore, hybrid models that merge CNNs with Transformers, such as BoTNet \cite{srinivas2021bottleneck}, UniFormer \cite{li2022uniformer}, and MaxViT \cite{tu2022maxvit}, have achieved success by combining the advantages of both designs. In the field of medical image analysis, customized ViT models have shown great promise. For example, the nnFormer integrates convolutional processes with self-attention for effective 3D medical image segmentation \cite{zhou2023nnformer}. The H2Former, fuses CNNs and Transformers for medical image segmentation, highlighting the efficacy of these models in specific applications \cite{he2023h2former}. Our study extends these innovations, emphasizing the collaborative benefits of CNNs and Transformers to improve feature representation in multi-phase medical image analysis.

\section{Methodology}
\label{method}
\subsection{Overview}
\begin{figure*}[h]
\centering
\includegraphics[scale=0.15]{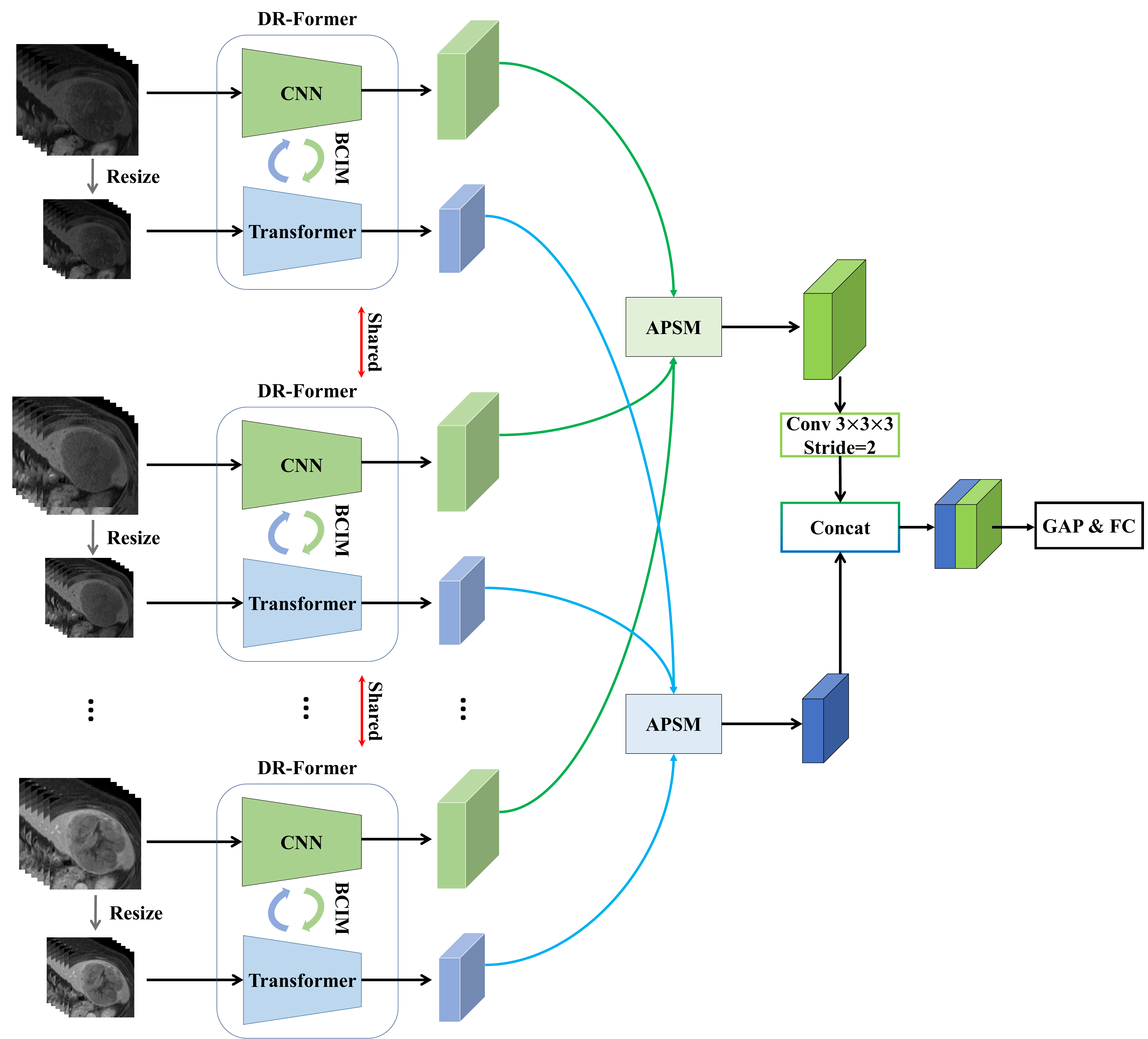}
\caption{
This diagram illustrates the SDR-Former framework, which utilizes a Siamese Neural Network (SNN) for processing each phase in the multi-phase imaging series independently yet simultaneously. Central to this framework is the hybrid Dual-Resolution Transformer (DR-Former), a fusion network that effectively merges CNN and Transformer approaches to handle high- and low-resolution inputs using shared weights. The Bilateral Cross-Resolution Integration Module (BCIM) is strategically incorporated into this architecture to facilitate semantic interactions and enhance the coalescence of the distinct high- and low-resolution streams. Furthermore, the Adaptive Phase Selection Module (APSM) plays a key role in dynamically merging these feature representations, ensuring a versatile and perceptive amalgamation of multi-phase features.
}
\label{SDR-Former}
\end{figure*}
As illustrated in Figure \ref{SDR-Former}, our SDR-Former framework employs a dual-stage strategy to effectively utilize multi-phase image data, consisting of an SNN for feature extraction and an APSM for integrating phase-specific features. 
\par
Unlike traditional multi-phase image analysis approaches (i.e., image-level and feature-level), the SDR-Former uniquely processes multi-phase images through a weight-sharing network of the SNN. Note that the number of weight-sharing networks employed in SNN within this study aligns with the number of input phases. This straightforward architecture ensures the model’s scalability and efficiency, as well as its seamless adaptation to datasets with different phase counts, thus enabling transfer learning across diverse multi-phase datasets. Nonetheless, the simple use of the SNN to extract multi-phase features may fall short in isolating distinctive phase features, potentially leading to weak feature representation. 
\par
To address this, we propose the novel DR-Former, a dual-branch network designed for the simultaneous processing of images at varying resolutions, serving as the weight-sharing network of the SNN. This configuration processes high-resolution images with a 3D CNN, adept at capturing spatial details, and low-resolution images with a 3D Transformer, aimed at recognizing global low-frequency contextual information. As a result, the weight-sharing mechanism intrinsic to the SNN underpins scalability, effectively accommodating datasets with diverse phase counts, thereby enabling transfer learning between such datasets. Moreover, incorporating the DR-Former within the SNN framework substantially enhances the model’s capability in representing features, particularly in multi-phase imaging scenarios as this dual approach allows the DR-Former to leverage the complementary strengths of CNNs and Transformers. Concurrently, the BCIM is introduced to bridge the CNN and Transformer pathways, facilitating a smooth exchange of features across different resolutions.
\par
Regarding post feature extraction, the APSM dynamically merges these features in a phase-sensitive fashion, highlighting the most diagnostically valuable features for each phase. The combined feature maps are then merged through a Global Average Pooling (GAP) layer and a Fully-Connected (FC) layer for final diagnostic classification. 
\par
\subsection{Dual-Resolution Transformer (DR-Former)}
Considering that high-resolution images naturally contain abundant high-frequency details, while low-resolution images predominantly focus on low-frequency elements \cite{zhang2018image,dai2019second}, our proposed DR-Former employs a CNN for high-resolution imagery and a Transformer for low-resolution imagery to optimize feature representation across different frequency ranges. 
\par
\begin{figure*}[h]
\centering
\includegraphics[width=0.8\textwidth]{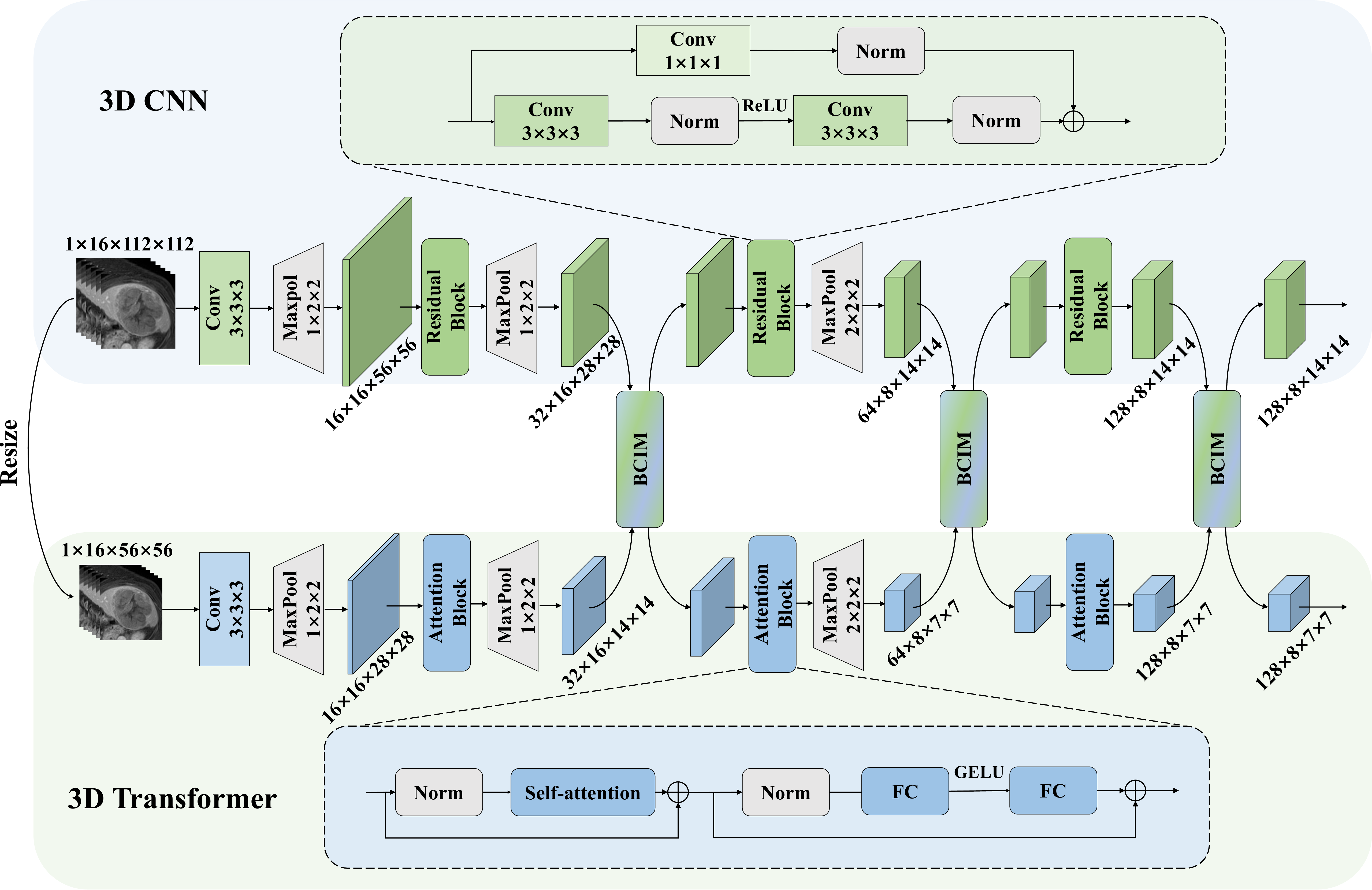}
\caption{
A diagrammatic depiction outlining the structure of the Hybrid Dual-Resolution Transformer (DR-Former).
}
\label{DRN}
\end{figure*}

As depicted in Figure \ref{DRN}, the DR-Former comprises two separate processing paths: a 3D CNN for analyzing high-resolution images and a 3D Transformer for low-resolution image encoding. The CNN pathway incorporates multiple residual blocks \cite{he2016deep}, with each block containing two $3 \times 3 \times 3$ convolutional layers linked by a shortcut connection. In parallel, the Transformer pathway consists of a sequence of attention blocks, each featuring a self-attention layer followed by a multi-layer perceptron (MLP). 
\par
Given the significant computational demands of 3D imaging and multi-phase data handling—particularly the quadratic complexity of the Multi-Head Self-Attention (MHSA) in the ViT \cite{dosovitskiy2021image}, we have evaluated four distinct self-attention schemes for computational efficiency: Shifted Window (Swin) \cite{liu2021swin}, Spatial Reduction Attention (SRA) \cite{wang2021pyramid}, Pooling-based Self-attention (PSA) \cite{wu2021p2t}, and Grid Self-attention (GSA) \cite{tu2022maxvit}. Our comprehensive experimental analyses, which will be detailed in Section \ref{sec:ab}, demonstrate GSA's superior ability to strike a balance between computational efficiency and performance.

\textbf{Bilateral Cross-resolution Integration Module (BCIM)} specifically designed to foster semantic consistency and elevate the feature interaction between high- and low-resolution data streams, as an effective and streamlined component, the BCIM facilitates the flexible information exchange between features processed by the CNN and Transformer paths. It enables both pathways to absorb relevant information from each other, thereby intensifying semantic links and enriching feature complementarity. This integration is crucial for the effective combination of high- and low-resolution features, thus boosting the overall efficacy of the DR-Former. 
\par
\begin{figure}[h]
\centering
\includegraphics[width=0.475\textwidth]{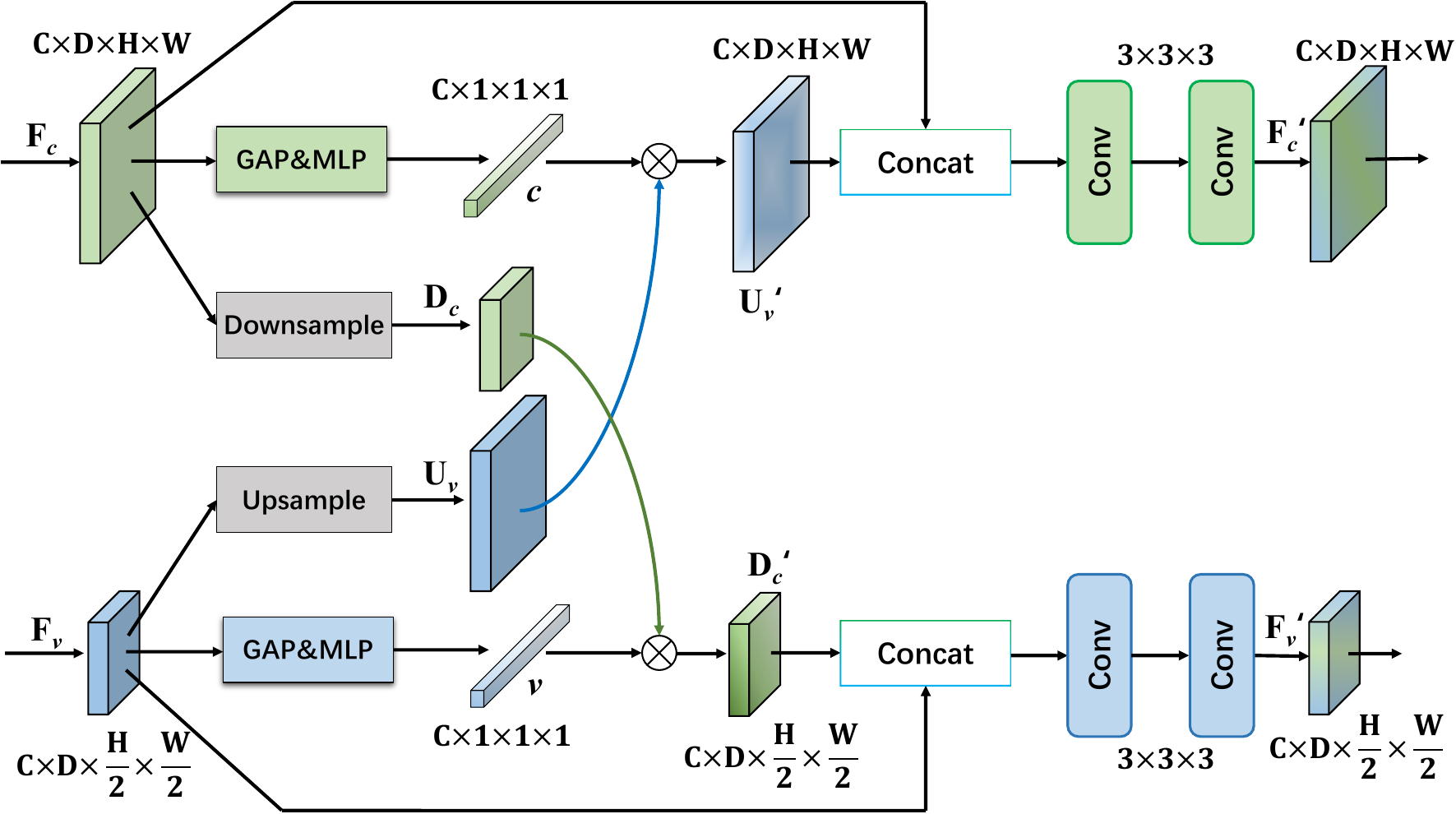}
\caption{A schematic illustration of the  Bilateral Cross-resolution Integration
 Module (BCIM).}
\label{BCIM}
\end{figure}

The mechanism of the BCIM is graphically illustrated in Figure \ref{BCIM}, demonstrating its role in relation to a high-resolution feature map $\textbf{F}{c} \in \mathbb{R} ^{C\times D\times H \times W }$ and a lower-resolution feature map $\textbf{F}{v} \in \mathbb{R} ^{C\times D\times \frac{H}{2} \times \frac{W}{2} }$, as encoded by their respective branches. Initially, GAP is applied to each feature map, denoted as $\left \{ {\textbf{F}{c}, \textbf{F}{v}} \right \}$. This step is followed by a two-layer Multi-Layer Perceptron (MLP) with a sigmoid activation function, yielding two vectors of attention coefficients $\left \{ \textbf{\emph{c}},\textbf{\emph{v}} \right \} \in \mathbb{R} ^{C\times1 \times1\times1 }$. Subsequently, $\textbf{F}{c}$ is down-scaled via average pooling to $\textbf{D}{c} \in \mathbb{R} ^{C\times D\times \frac{H}{2} \times \frac{W}{2}}$, while $\textbf{F}{v}$ is up-scaled to $\textbf{U}{v} \in \mathbb{R} ^{C\times D\times H \times W }$. The subsequent step involves the element-wise multiplication of $\textbf{U}{v}$ with \textbf{\emph{c}} and $\textbf{D}{c}$ with \textbf{\emph{v}}, producing modified feature maps $\left \{ {\mathbf{U}{v}' \in \mathbb{R} ^{C\times D \times H\times W }, \mathbf{D}{c}' \in \mathbb{R} ^{C\times D \times \frac{H}{2}\times \frac{W}{2} }} \right \}$ aimed at enabling semantic integration. These adjusted maps $\left \{ {\mathbf{U}{v}', \mathbf{D}{c}' }\right \}$ are then merged with their original feature maps $\left \{ {\textbf{F}{c}, \textbf{F}{v}} \right \}$. The combined maps are then processed through two $3 \times 3 \times 3$ convolutional layers, each followed by Batch Normalization (BN) and ReLU activation, resulting in the finalized output feature maps $\left \{ {\textbf{F}{c}', \textbf{F}{v}'} \right \}$. The BCIM can be formally described as follows:
\begin{equation}
\begin{array}{c}
c  =  \sigma(MLP(\sum_{i  =  1}^{D} \sum_{j  =  1}^{H} \sum_{k  =  1}^{W} \mathbf{F}_{c}(i,j,k))) \\
v  =  \sigma(MLP(\sum_{i  =  1}^{D} \sum_{j  =  1}^{H} \sum_{k  =  1}^{W} \mathbf{F}_{v}(i,j,k))) \\
\mathbf{D}_{c}  =  Down(\mathbf{F}_{c}); \mathbf{U}_{v} = Up(\mathbf{F}_{v}) \\
\mathbf{D}_{c}'  = v \times \mathbf{D}_{c}; \mathbf{U}_{v}'  = c \times \mathbf{U}_{v} \\
\mathbf{F}_{c}'  = W_{3}^{2}( \hat{\mathcal{C} } (\mathbf{F}_{c},\mathbf{U}_{v}')); \mathbf{F}_{v}' = W_{3}^{2}( \hat{\mathcal{C} } (\mathbf{F}_{v},\mathbf{D}_{c}')) \\
\end{array}
\end{equation}
Here, the terms $\sigma(\cdot)$, $\hat{\mathcal{C}}(\cdot)$, and $W_{3}^{2}(\cdot)$ denote the sigmoid activation function, the operation of channel-wise concatenation, and a sequence of two $3 \times 3 \times 3$ convolutional layers, each followed by BN and ReLU activation, respectively.

\par

\subsection{Adaptive Phase Selection Module (APSM)}
Inspired by the critical role of phase-specific imaging attributes in clinical diagnostics \cite{joo2020radiologic,xiao2015multiple}, we have developed the APSM, tailored to facilitate detailed interaction among phases and dynamically modulate the focus on features from each phase in line with their clinical significance. 

\begin{figure}[h]
\centering
\includegraphics[width=0.485\textwidth]{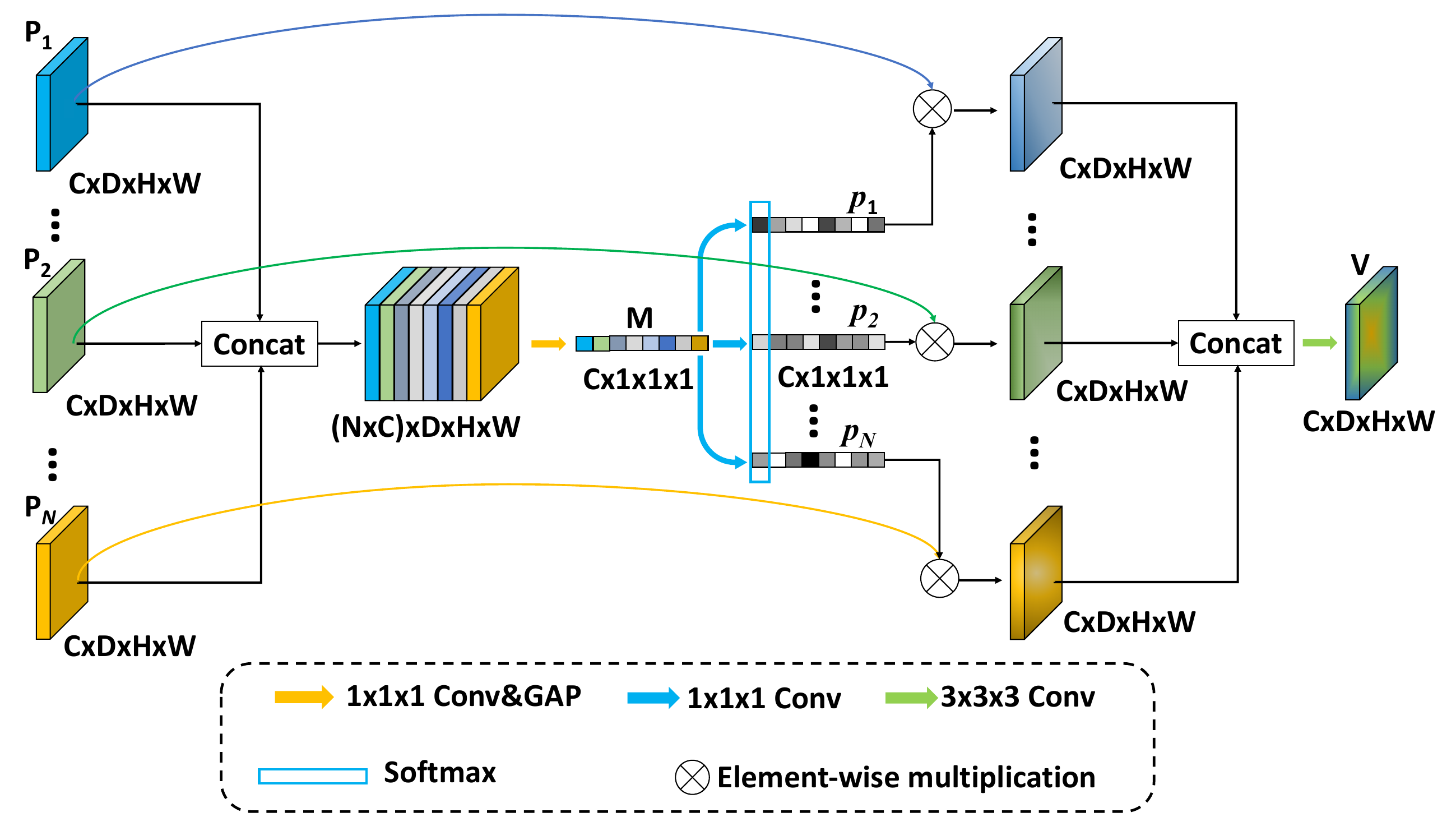}
\caption{
A schematic illustration of the Adaptive Phase Selection Module (APSM).
}
\label{APSM}
\end{figure}

As illustrated in Figure \ref{APSM}, the APSM processes feature maps for each phase, $\left \{ \mathbf{P}_{1}, \mathbf{P}_{2}, \cdots ,\mathbf{P}_{N} \right \} \in \mathbb{R} ^{C\times D\times H\times W}$, generated by the DR-Former, where \emph{N} indicates the total number of phases. The process initiates with a channel-wise concatenation of the multi-phase feature maps, followed by refinement using a $1\times1\times1$ convolution and a GAP, leading to a unified global descriptor $\mathbf{M} \in \mathbb{R} ^{C\times 1\times 1\times 1}$. Following this, \emph{N} parallel $1\times1\times1$ convolutions are applied to create individual global descriptors $\left \{ \mathbf{M}_{1}, \mathbf{M}_{2}, \cdots ,\mathbf{M}_{N} \right \}$. These descriptors are then used to calculate phase-specific attention coefficients $\left \{ p{1}, p{2}, \cdots ,p{N} \right \}$ via a channel-wise softmax function, ensuring that the impact of each phase is proportionally weighted according to its diagnostic relevance, mathematically expressed as:
\begin{equation}
   p_{k} = \frac{{e^{\mathbf{M} _{k}^{j}  } } }{e^{\sum_{i = 1}^{N}\mathbf{M}_{i}^{j}} } 
\end{equation}
Here, \emph{j} = $\left \{ 1, 2, 3, \cdots , C \right \} $ and \emph{k} = $\left \{ 1, 2, 3, \cdots , N \right \} $. The attention-weighted feature maps are concatenated and further refined by a $3\times3\times3$ convolution with BN and ReLU activation to produce the final feature map $\mathbf{V}\in \mathbb{R} ^{C\times D\times H\times W}$ as expressed by the following equation:
\begin{equation}
\mathbf{V} = W_{3} (\hat{\mathcal{C}}(p_{1} \times \mathbf{P}{1}, p{2} \times \mathbf{P}{2}, \cdots ,p{N} \times \mathbf{P}_{N}))
\end{equation}
\par
In the architecture of the SDR-Former, the APSM is tactically situated following the encoding of features from both high- and low-resolution images across multiple phases. This placement empowers the model to adaptively regulate the significance of features across different resolutions. As a result, the diagnostic accuracy of the SDR-Former is notably improved, owing to its proficient integration of phase-specific and inter-phase contextual information.

\section{Experimental Configuration}
\label{experiment_set}
\subsection{Datasets}
In this study, the performance of the SDR-Former is assessed using two distinct multi-phase imaging datasets derived from clinical sources: one dataset comprises CT images, while the other includes MR images.
\par
\textbf{CT Image Dataset:} 
This dataset encompasses 4,304 multi-phase liver lesions from 10 different hospitals in China, involving over 2,000 patients. Each lesion is documented in three separate phases: venous, arterial, and non-contrast, forming distinct volumetric data. We implement patient-level segregation to split the dataset into training, validation, and test groups, ensuring no data overlap patient-wise. The training group contains 3,248 lesions (1,625 benign, 1,623 malignant), the validation group includes 434 lesions (253 benign, 181 malignant), and the testing group comprises 622 lesions (339 benign, 283 malignant), all of which are from an independent hospital for testing purpose.
\par
\textbf{MR Image Dataset:} 
Sourced from Ningbo Medical Center Lihuili Hospital, this dataset includes 498 annotated multi-phase liver lesions from an equal number of patients. Each lesion is represented across eight phases: non-contrast, arterial, venous, delay, T2-weighted imaging, diffusion-weighted imaging, T1 in-phase, and T1 out-of-phase, each contributing an individual volume. The lesions are classified into seven categories: hepatocellular carcinoma (HCC), intrahepatic cholangiocarcinoma (ICC), liver metastases (HM), hepatic cysts (HC), hepatic hemangioma (HH), focal nodular hyperplasia (FNH), and hepatic abscess (HA). Pre-partitioned for research purposes, the dataset is divided into training (316 lesions), validation (78 lesions), and testing sets (104 lesions). In line with our commitment to advancing multi-phase medical imaging analysis, this dataset has been publicly released. Importantly, the MICCAI LLD-MMRI Challenge\footnote{https://github.com/LMMMEng/LLD-MMRI2023} utilizes this dataset as its core resource, aiming to promote the development and refinement of computer-aided diagnosis (CAD) systems within this field. Furthermore, within the scope of this paper, we provide annotations for an additional 104 cases, constituting a testing set that is not incorporated within the challenge. 
\subsection{Implementation Details}
Our SDR-Former framework is developed using the PyTorch framework and all experiments are conducted on a NVIDIA RTX 3090 GPU. For training, we employ the AdamW optimizer, starting with an initial learning rate of 0.0001, which is dynamically adjusted using a cosine annealing schedule to fine-tune performance. A weight decay of 0.05 is applied to prevent overfitting. The training process spans over 200 epochs, with the first 5 epochs designated as a warm-up phase where the learning rate gradually escalates. The standard cross-entropy loss function is employed to compare the model's output with the ground-truth. Due to hardware limitations, our batch size is set to 8, and each lesion volume is uniformly resized to $16\times128\times128$. To further mitigate overfitting, a variety of data augmentation techniques are employed, including random rotations, erasing, and flips across different anatomical axes. In the training stage, we randomly crop the lesion volumes to $14\times112\times112$, while for evaluation, a central crop of the same size is extracted. The model's effectiveness is rigorously assessed using a range of established metrics: Accuracy (ACC), Area Under the Curve (AUC), F1-score (F1), and Cohen's Kappa (Kappa). We also conduct a paired t-test to statistically ascertain and differentiate the performance of our model against benchmark methods, ensuring an in-depth and statistically validated performance analysis.

\section{Results and Discussion}
\label{results}
This section delves into a thorough assessment of the SDR-Former. First, an ablation study is conducted to determine the structural effectiveness of the hybrid DR-Former architecture via comparing it with traditional CNN and Transformer models. Following this, the specific roles of the BCIM and the APSM in the model are meticulously examined, emphasizing their influence on the model's performance. The analysis further encompasses various self-attention mechanisms embedded in the model's Transformer branch, assessing their role in augmenting feature extraction efficiency. Additionally, the importance of utilizing multi-phase imaging is investigated. A detailed comparison with the current state-of-the-art methods is presented, establishing the proposed SDR-Former as a new benchmark in the field. The final part of the evaluation explores the model's ability for cross-modality transfer learning and its adaptability to different datasets, affirming the SDR-Former's flexibility and scalability in medical image analysis.

\subsection{Comparison with State-of-the-Art (SOTA) Methods}
\label{sec:sota_compare}
\begin{figure}[h]
\centering
\includegraphics[scale=0.6]{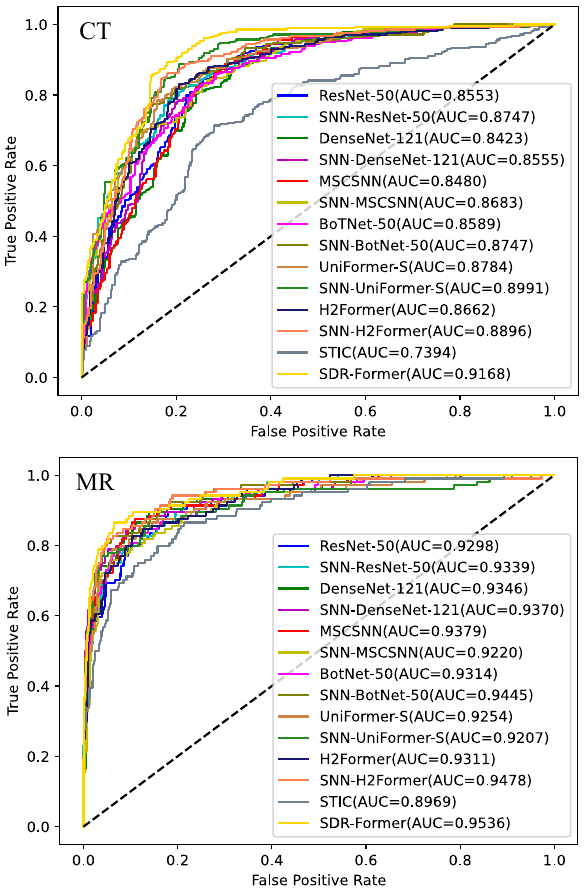}
\caption{
Receiver Operating Characteristic (ROC) curves for various methods.
}
\label{SOTA_compare_ROC}
\end{figure}

In this work, the performance of the SDR-Former model is benchmarked against a suite of SOTA models. This comparison includes two renowned CNN architectures, ResNet-50 \cite{he2016deep} and DenseNet-121 \cite{huang2017densely}, along with two advanced Transformer-based models, BoTNet-50 \cite{srinivas2021bottleneck} and UniFormer-S \cite{li2022uniformer}. To highlight the advantages of our hybrid design, we also included the MSCSNN \cite{xu2020mscs}, designed for multi-resolution input, and the H2Former \cite{he2023h2former}, another hybrid model. Note that H2Former is principally intended for segmentation, so we employed only its encoder in our experiments. Additionally, we explored the resilience of our multi-phase analysis approach by integrating these models into the SDR-Former's weight-sharing network framework, resulting in variants like SNN-ResNet-50, SNN-DenseNet-121, SNN-BoTNet-50, SNN-UniFormer-S, SNN-MSCSNN, and SNN-H2Former, all under the umbrella of an SNN architecture. For a thorough and unbiased evaluation, we also considered STIC \cite{gao2021deep}, a publicly well-known method for multi-phase image classification. The outcomes of this comprehensive comparative study are summarized in Table \ref{tab:compare} and depicted in the ROC curves shown in Figure \ref{SOTA_compare_ROC}.

\begin{table}[htbp]
  \centering
\caption{Comparative Performance Metrics for Multi-phase Lesion Classification Across CT and MR Datasets. }
\label{tab:compare}
    \begin{tabular}{lllll}
    \toprule
    Method & ACC   & AUC   & F1    & Kappa \\
    \midrule
    \textbf{CT (3-phase)} &       &       &       &  \\
    STIC  & 0.6913  & 0.7394  & 0.6933  & 0.3883  \\
    ResNet-50 & 0.7814  & 0.8553  & 0.7834  & 0.5670  \\
    SNN-ResNet-50 & 0.7974  & 0.8747  & 0.7872  & 0.5946  \\
    DenseNet-121 & 0.7588  & 0.8423  & 0.7611  & 0.5224  \\
    SNN-DenseNet-121 & 0.7669  & 0.8555  & 0.7745  & 0.5403  \\
    MCSCNN & 0.7749  & 0.8480  & 0.7603  & 0.5485  \\
    SNN-MCSCNN & 0.7621  & 0.8683  & 0.7716  & 0.5314  \\
    BoTNet-50 & 0.7669  & 0.8589  & 0.7724  & 0.5395  \\
    SNN-BoTNet-50 & 0.8087  & 0.8747  & 0.8072  & 0.6143  \\
    UniFormer-S & 0.7862  & 0.8784  & 0.7932  & 0.5783  \\
    SNN-UniFormer-S & 0.8312  & \underline{0.8991}  & 0.8305  & 0.6611  \\
    H2Former & 0.8103  & 0.8662  & 0.8097  & 0.6200  \\
    SNN-H2Former & \underline{0.8344}  & 0.8896  & \underline{0.8335}  & \underline{0.6681}  \\
    SDR-Former & \textbf{0.8537 } & \textbf{0.9168 } & \textbf{0.8529 } & \textbf{0.7058 } \\
    \textit{p-value} & \textless 0.01 & \textless 0.01 & \textless 0.01 & \textless 0.01 \\
    
     &       &       &       &  \\
    \textbf{MR (8-phase)} &       &       &       &  \\
    STIC  & 0.6346  & 0.8969  & 0.6254  & 0.5572  \\
    ResNet-50 & 0.6923  & 0.9298  & 0.6898  & 0.6244  \\
    SNN-ResNet-50 & 0.7115  & 0.9339  & 0.7168  & 0.6541  \\
    DenseNet-121 & 0.7404  & 0.9346  & 0.7171  & 0.6797  \\
    SNN-DenseNet-121 & 0.7500  & 0.9370  & 0.7394  & 0.6961  \\
    MCSCNN & 0.7115  & 0.9379  & 0.7089  & 0.6536  \\
    SNN-MCSCNN & 0.7308  & 0.9220  & 0.7409  & 0.6770  \\
    BoTNet-50 & 0.7212  & 0.9314  & 0.7139  & 0.6628  \\
    SNN-BoTNet-50 & 0.7692  & 0.9445  & 0.7572  & 0.7135  \\
    UniFormer-S & 0.7308  & 0.9254  & 0.7123  & 0.6662  \\
    SNN-UniFormer-S & 0.7692  & 0.9207  & 0.7639  & 0.7250  \\
    H2Former & 0.7212  & 0.9311  & 0.7342  & 0.6660  \\
    SNN-H2Former & \underline{0.7788}  & \underline{0.9478}  & \underline{0.7745}  & \underline{0.7263}  \\
    SDR-Former & \textbf{0.7885 } & \textbf{0.9536 } & \textbf{0.7910 } & \textbf{0.7467 } \\
    \textit{p-value} & \textless 0.05 & \textless 0.01 & \textless 0.01 & \textless 0.01 \\
    \bottomrule
    \end{tabular}%
\end{table}%
\par
Firstly, the STIC method exhibits the least favorable performance among the methods evaluated, potentially due to its naive fusion of CNN and RNN, which may not be adequate for fully capturing the intricate relationships present in multi-phase images. Secondly, our hybrid CNN-Transformer model significantly outperforms the MSCSNN, which utilizes three parallel ResNets for multi-resolution data processing. This highlights the superior ability of our model to handle such complex inputs effectively. Thirdly, our model exceeds the performance of the H2Former, probably due to our deliberate allocation of different resolutions to the specific CNN and Transformer branches. This method fosters a more nuanced interaction of features compared to H2Former's simpler aggregate fusion approach. Notably, traditional CNN-based models, such as SNN-DenseNet, SNN-ResNet, and SNN-MCSCNN, exhibit only modest improvements within our multi-phase framework, indicating limited benefits over their single-phase versions. On the contrary, Transformer-based models demonstrate more substantial performance enhancements in the multi-phase context, owing to their inherent dynamic self-attention capabilities. These capabilities enable Transformer models to generate phase-sensitive attention maps, a functionality not inherent in CNN-exclusive models. Ultimately, the integration of Transformers within our multi-phase framework substantially elevates performance, underscoring the importance of Transformer-based self-attention in multi-phase lesion classification tasks.
\par
Overall, the proposed SDR-Former model exhibits outstanding performance compared to other SOTA models, as evidenced by its robust performance metrics. In a CT dataset with three phases and two lesion types, the SDR-Former model achieves an ACC of {0.8537}, an AUC of {0.9168}, an F1 of {0.8529}, and a Kappa {0.7058}. In a more complex MR dataset featuring eight phases and seven lesion categories, the model obtains an ACC of {0.7885}, an AUC of {0.9536}, an F1 of {0.7910}, and a Kappa of {0.7467}. Furthermore, to establish statistical significance, we conducted independent two-sample t-tests comparing the performance of our SDR-Former with the best-performing method in each metric. The null hypothesis assumes that our method doesn't surpass the best-performing method. Our analysis reveals p-values below 0.01 (or 0.05) for SDR-Former across both datasets, indicating compelling evidence against the null hypothesis. Consequently, our approach demonstrates significant advancements over previous methods in two distinct datasets, highlighting its superiority.
\subsection{Ablation Study}
\label{sec:ab}
\subsubsection{Effectiveness of the DR-Former} 
The DR-Former, central to the feature extraction process of the SDR-Former framework, is constructed as a hybrid architecture combining CNN and Transformer mechanisms. This design's foundation is the exploitation of the collaborative strengths of convolutional operations and self-attention mechanisms \cite{park2022vision}. 
\par
In our implementation, high-resolution images are processed through the DR-Former's convolutional branch, while its self-attention branch handles the low-resolution images. This strategic allocation is designed to fully utilize the unique benefits of each processing approach. To confirm the DR-Former's hybrid design's efficiency, we conducted a performance analysis of its individual CNN and Transformer branches, applying each separately for feature extraction. The results of this analysis are detailed in Table \ref{tab:Subnetworks_compare}.

\begin{table}[h]
  \centering
  \caption{Performance Assessment of Individual Sub-Networks in the Hybrid Dual-Resolution Transformer (DR-Former).}
  \scalebox{1.0}{
    \begin{tabular}{lllll}
    \toprule
    Method & ACC   & AUC   & F1    & Kappa \\
    \midrule
    \textbf{CT  (3-phase)} &       &       &       &  \\
    Only Transformer & 0.7621  & 0.8552  & 0.7549  & 0.5131  \\
    Only CNN & 0.7765  & 0.8733  & 0.7705  & 0.5434  \\
    SDR-Former & \textbf{0.8537 } & \textbf{0.9168 } & \textbf{0.8529 } & \textbf{0.7058 } \\
          &       &       &       &  \\
    \textbf{MR (8-phase)} &       &       &       &  \\
    Only Transformer & 0.6731  & 0.9360  & 0.6721  & 0.6056  \\
    Only CNN & 0.6827  & 0.9228  & 0.7078  & 0.6282  \\
    SDR-Former & \textbf{0.7885 } & \textbf{0.9536 } & \textbf{0.7910 } & \textbf{0.7467 } \\
    \bottomrule
    \end{tabular}%
    }
  \label{tab:Subnetworks_compare}%
\end{table}%
\par
The experimental results robustly support the notion that exclusive reliance on either CNN or Transformer architectures separately results in less than optimal performance. Notably, the standalone usages of CNNs demonstrate superior outcomes compared to Transformers used alone. This underscores potential challenges in optimizing pure Transformer models, as suggested in recent research \cite{xiao2021early}. The DR-Former's design, which merges CNN and Transformer and seamlessly integrates them through the BCIM, achieves notably enhanced performance over the singular use of each technique. This dual-branch, synergistic method excels in extracting multi-resolution features. The CNN's inherent biases aid in more efficient training of the Transformer component, leading to a notable increase in the precision of lesion classification. 

\subsubsection{Influence of Core Integral Modules} 
The BCIM and APSM are crucial components of our SDR-Former architecture, each bringing unique enhancements to the processing of multi-phase features. The BCIM boosts the feature representation by facilitating interactions between convolutional and self-attention-derived features within each imaging phase. Simultaneously, the APSM is designed for dynamic inter-phase communication, enhancing the significance of features across phases based on the particular characteristics of the input lesions. To thoroughly assess the individual and collective contributions of these components, a set of ablation studies was performed on both CT and MR datasets. Initially, we established a Baseline model devoid of the contextual interaction offered by the BCIM and the APSM. This Baseline served as a comparative foundation for further experiments. Subsequently, the Baseline was enhanced with the APSM module, termed 'w/ APSM,' to gauge its effect on the adaptive prioritization of phase-specific features. In a separate experiment, the BCIM module was integrated into the Baseline, labeled 'w/ BCIM,' to evaluate the benefit of improved feature interaction. The final configuration, encompassing both the APSM and BCIM, was analyzed to determine their combined influence within the SDR-Former framework. The results from these analyses are concisely depicted in Table \ref{tab:att_compare}.
\begin{table}[htbp]
  \centering
  \caption{Evaluation of the Effects of Integrating Various Network Components.}
  \scalebox{1.0}{
    \begin{tabular}{lllll}
    \toprule
    Method & ACC   & AUC   & F1    & Kappa \\
    \midrule
    \textbf{CT  (3-phase)} &       &       &       &  \\
    Baseline & 0.7974  & 0.8776  & 0.7942  & 0.5889  \\
    Baseline+BCIM & 0.8376  & 0.9134  & 0.8359  & 0.6720  \\
    Baseline+APSM & 0.8055  & 0.8995  & 0.8018  & 0.6044  \\
    SDR-Former & \textbf{0.8537 } & \textbf{0.9168 } & \textbf{0.8529 } & \textbf{0.7058 } \\
          &       &       &       &  \\
    \textbf{MR (8-phase)} &       &       &       &  \\
    Baseline & 0.7404  & 0.9410  & 0.7508  & 0.6908  \\
    Baseline+BCIM & 0.7692  & 0.9418  & 0.7730  & 0.7196  \\
    Baseline+APSM & 0.7596  & 0.9520  & 0.7636  & 0.7051  \\
    SDR-Former & \textbf{0.7885 } & \textbf{0.9536 } & \textbf{0.7910 } & \textbf{0.7467 } \\
    \bottomrule
    \end{tabular}%
    }%
  \label{tab:att_compare}%
\end{table}%
\par
The findings offer compelling evidence of the performance enhancements attributed to the BCIM and APSM modules within the SDR-Former model, as evidenced by various quantitative metrics. Notably, the APSM module demonstrates a more pronounced impact on the MR dataset compared to the CT dataset. This likely stems from the MR dataset's greater phase variety, encompassing eight phases versus the CT dataset's three, enabling the APSM to more effectively perform adaptive feature selection across a broader range of phase scenarios. Regarding the BCIM, an observable boost in model performance is noted on both datasets, underscoring the module's vital role in mediating contextual information exchange between the CNN and Transformer branches, thus improving feature representation. The simultaneous incorporation of BCIM and APSM leads to even more substantial improvements, verifying the advantages of their synergistic integration as reflected in the outcomes. Furthermore, Table \ref{tab:att_compare_params} provides insights into the computational load associated with BCIM and APSM. The results indicate that despite their relatively low computational requirements, these modules make a significant contribution to overall performance, confirming their effectiveness and efficiency.
\begin{table}[h]
  \centering
  \caption{Evaluation of Computational Complexity for Various Models: Benchmarks Performed Using a Single NVIDIA GeForce RTX 3090 GPU with Multi-Phase Inputs. Dimensions: 14 $\times$ 112 $\times$ 112.}
  \scalebox{1}{
    \begin{tabular}{rlcc}
    \toprule
    \multicolumn{1}{l}{\textit{N}} & Method & FLOPs (G) & Params (M) \\
    \midrule
    \multicolumn{1}{l}{\textbf{3 Phases} }& Baseline & 102.07  & 27.87  \\
          & w/ BCIM & 102.29  & 28.43  \\
          & w/ APSM & 102.07  & 27.96  \\
          & SDR-Former & 102.30  & 28.52  \\
          &       &       &  \\
    \multicolumn{1}{l}{\textbf{8 Phases}} & Baseline & 40.17  & 19.02  \\
          & w/ BCIM & 40.26  & 19.25  \\
          & w/ APSM & 40.18  & 19.11  \\
          & SDR-Former & 40.26  & 19.34  \\
    \bottomrule
    \end{tabular}}
  \label{tab:att_compare_params}%
\end{table}%

\textbf{Visualization of BCIM's Role:} The Grad-CAM technique \cite{selvaraju2020grad} was employed to visualize the impact of the BCIM on activation patterns within the SDR-Former model. This visualization targets intermediate output feature maps from both CNN and Transformer branches. 

\begin{figure}[htpb]
\centering
\includegraphics[scale=0.175]{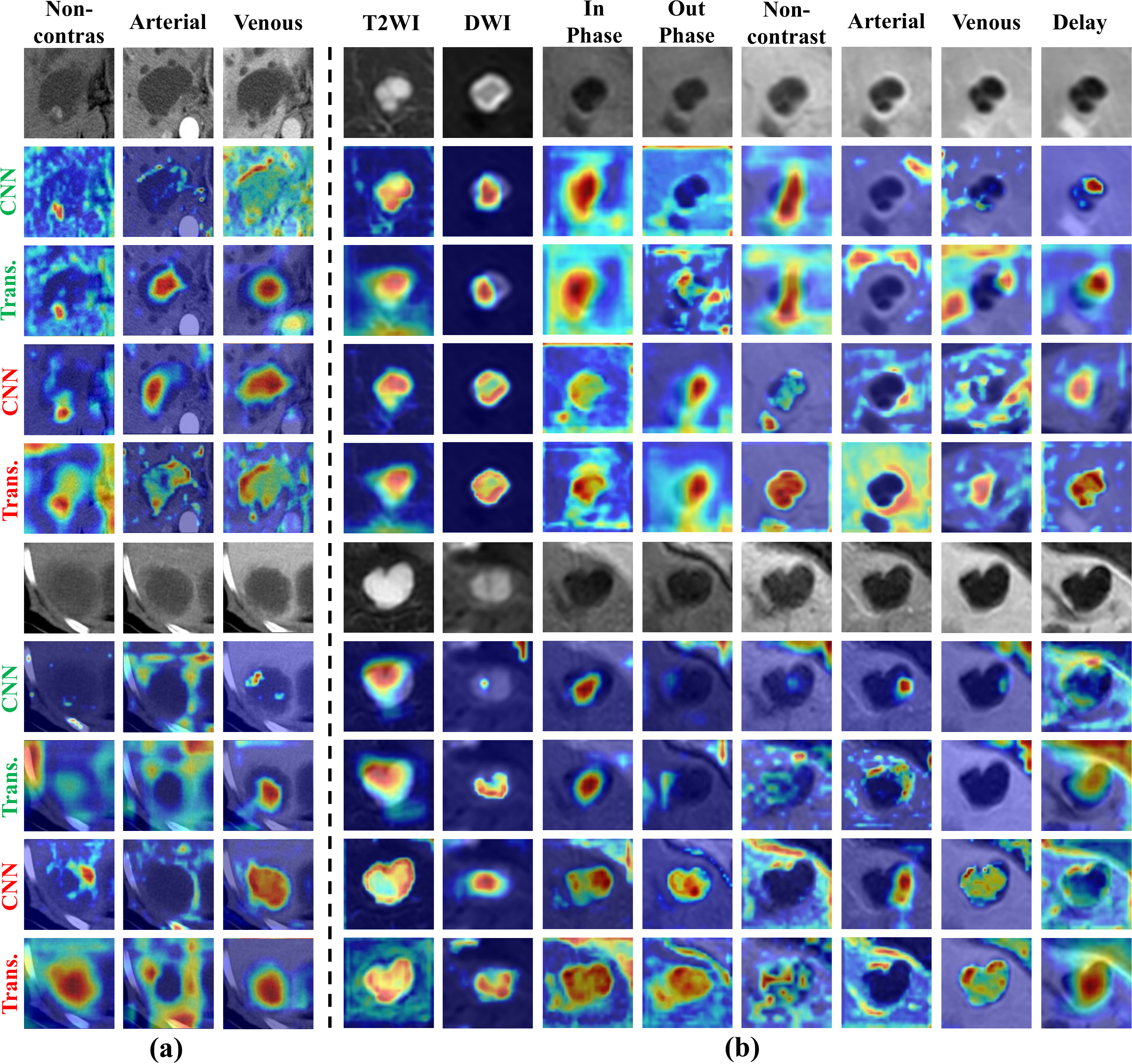}
\caption{Feature Activation Visualization Using Grad-CAM: Comparative Analysis for (a) CT and (b) MR Datasets. Green overlays indicate the model without BCIM, whereas red overlays depict the model with BCIM integrated. The abbreviations `CNN' and `Trans.' refer to the CNN and Transformer segments of the SDR-Former model, respectively. For enhanced detail, zooming in is advisable.}
\label{vis_bcim}
\end{figure}

Figure \ref{vis_bcim} illustrates that, in configurations without the BCIM, the CNN branch exhibits relatively limited receptive fields, while the Transformer branch shows numerous non-specific activations. However, with the inclusion of the BCIM, which encourages interaction between the branches, a notable improvement in feature representation quality is observed. This enhanced representation leads to more precise lesion localization in various imaging phases. The CNN branch becomes more efficient in pinpointing localized features without excess noise, and the Transformer branch, utilizing its capacity for global context comprehension, accurately outlines lesion contours. To conclude, the BCIM acts as a connective mechanism, effectively reducing the disparity between the two branches and enabling them to work together more effectively, thus enriching the quality of multi-phase feature representations.

\begin{figure}[htpb]
\centering
\includegraphics[scale=0.175]{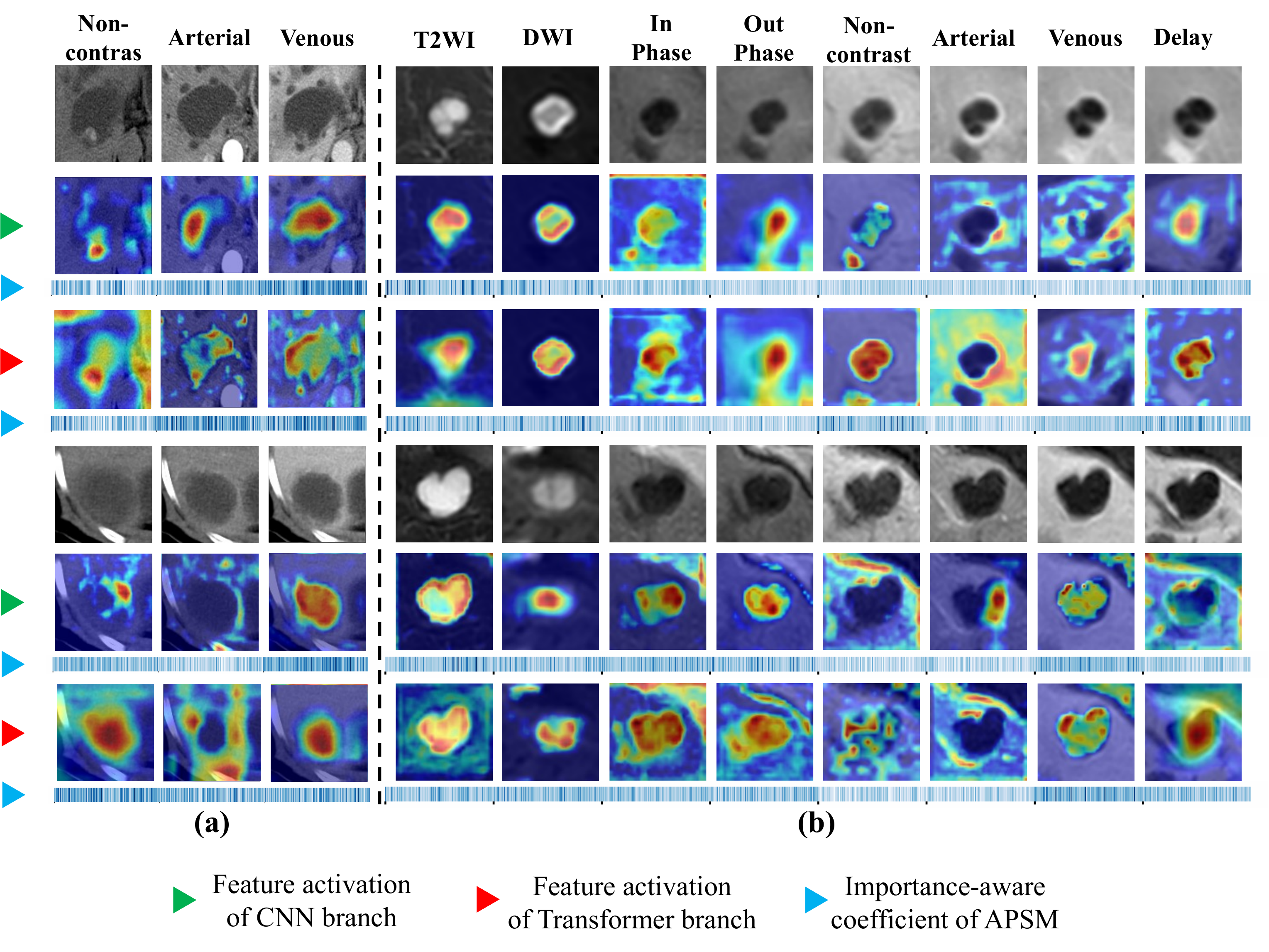}
\caption{Illustration of Significance-Oriented Coefficients Produced by the Adaptive Phase Selection Module (APSM) for (a) CT and (b) MR Datasets. The intensity of the blue color indicates the strength of the coefficients, reflecting the impact of the APSM on the selection of phase-specific features. For enhanced detail, zooming in is advisable.}
\label{vis_apsm}
\end{figure}

\textbf{Visualization of APSM's Role:} The adaptive phase selection capability of the APSM is also visually illustrated through its `Attention Coefficients' mapping. This visualization aims to exhibit how APSM variably allocates attention weights across phases, based on the relative importance of their feature activations. As shown in Figure \ref{vis_apsm}, the APSM is adept at differentiating between phases, assigning greater attention to those with more pronounced feature activations that suggest the presence of lesions. Additionally, the attention variation among phases validates the module's effectiveness in identifying the most informative phases for precise lesion localization. This confirms APSM's dynamic phase selection capacity and its significant contribution to the enhanced performance of the SDR-Former model.

\subsubsection{Selection of Self-attention Modules}
In our investigation to determine the most suitable self-attention architecture for the Transformer branch of our framework, we conducted an analytical comparison of four contemporary self-attention designs: Swin \cite{liu2021swin}, SRA \cite{wang2021pyramid}, PSA \cite{wu2021p2t}, and GSA \cite{tu2022maxvit}. Originally developed for 2D imagery, these architectures underwent necessary adaptations for their application in 3D imaging scenarios. These modifications were meticulously implemented following the principles of the original methodologies, ensuring a smooth transition from 2D to 3D image processing capabilities within our model.

\begin{table}[htpb]
  \centering
  \caption{Performance Analysis with Various Self-Attention Mechanisms.}
  \scalebox{1}{
  \setlength{\tabcolsep}{1mm}{
    \begin{tabular}{lllll}
    \toprule
    Method & ACC   & AUC   & F1    & Kappa \\
    \midrule
    \textbf{CT  (3-phase)} &       &       &       &  \\
    Swin  & 0.8087  & 0.8989  & 0.8086  & 0.6221  \\
    SRA   & 0.8264  & 0.9127  & 0.8230  & 0.6468  \\
    PSA   & 0.8441  & 0.9075  & 0.8434  & 0.6871  \\
    GSA   & \textbf{0.8537 } & \textbf{0.9168 } & \textbf{0.8529 } & \textbf{0.7058 } \\
          &       &       &       &  \\
    \textbf{MR (8-phase)} &       &       &       &  \\
    Swin  & 0.7787  & 0.9508  & 0.7738  & 0.7511  \\
    SRA   & 0.7788  & 0.9530  & 0.7760  & 0.7705  \\
    PSA   & 0.7884  & 0.9507  & 0.7830  & 0.7700  \\
    GSA   & \textbf{0.7885 } & \textbf{0.9536 } & \textbf{0.7910 } & \textbf{0.7467 } \\
    \bottomrule
    \end{tabular}%
    }}
  \label{tab:VIT_compare}%
\end{table}
The comparative analysis of self-attention modules, as outlined in Table \ref{tab:VIT_compare}, indicates that Swin's efficacy is somewhat lower compared to other models. The architecture of Swin, with its emphasis on localized window-based attention, may not be as effective in capturing expansive relationships, a key element in Transformer design. Conversely, SRA and PSA maintain a more comprehensive view of self-attention by evaluating token-to-region relationships, though they do reduce the dimensions of key and value matrices. While this approach betters Swin's output, it comes with the potential shortcoming of information diminution due to feature downsampling. GSA, however, stands out as the most effective among the evaluated modules. It retains the full architecture of the query, key, and value matrices, preventing any unintended loss of information. Additionally, GSA employs a sparse token-linking strategy that broadens its range to a global scale, significantly boosting its utility in the evaluated tasks.

\subsubsection{Benefits of Multi-Phase Imaging in Diagnostic Accuracy}
To evaluate the impact of multi-phase imaging on enhancing diagnostic accuracy, we performed an ablation study using the CT dataset. It is important to note that with single-phase input, the APSM is not utilized. Table \ref{tab:missing_phase} demonstrates that relying solely on non-contrast phases leads to the least favorable outcomes, aligning with clinical observations. The contrast-enhanced phases significantly improve diagnostic performance as arterial and/or venous phases enhance lesion visibility. Especially noteworthy is the venous phase, which excels in clearly differentiating lesions from adjacent tissues. Our analysis also highlights the SDR-Former’s capability to effectively utilize multi-phase data, evidently outperforming single-phase approaches. This substantiates the premise that multi-phase data is vital in enriching feature depiction for more precise lesion identification. The model achieves optimal performance when incorporating all available phases, thereby emphasizing the critical importance of a comprehensive multi-phase imaging approach for accurate liver lesion classification.
\par
\begin{table}[h]
  \centering
  \caption{Influence of Multi-Phase Imaging in Diagnostic Accuracy.}
  \scalebox{0.75}{
    \begin{tabular}{cccllll}
    \toprule
    Non-contrast & Arterial & Venous & ACC   & AUC   & F1    & Kappa \\
    \midrule
    \checkmark     &       &       & 0.7788  & 0.8594  & 0.7787  & 0.5606  \\
          & \checkmark     &       & 0.7885  & 0.8516  & 0.7865  & 0.5736  \\
          &       & \checkmark    & 0.7981  & 0.8241  & 0.7938  & 0.5907  \\
    \checkmark     & \checkmark     &       & 0.8077  & 0.8627  & 0.8051  & 0.6115  \\
    \checkmark     &       & \checkmark     & 0.8173  & 0.8579  & 0.8159  & 0.6322  \\
          & \checkmark     & \checkmark     & 0.8269  & 0.8675  & 0.8246  & 0.6504  \\
    \checkmark     & \checkmark     & \checkmark     & \textbf{0.8537} & \textbf{0.9168 } & \textbf{0.8529} & \textbf{0.7058 } \\
    \bottomrule
    \end{tabular}%
    }
  \label{tab:missing_phase}%
\end{table}%
\par
\subsection{Efficacy in Cross-Modality Transfer Learning}
Our proposed SDR-Former model is inherently designed for effective transfer learning across datasets with varied phase counts, showcasing its scalability. To validate this capability, we carried out a set of empirical tests. Initially, the SDR-Former was trained using our three-phase CT dataset, then its learned weights were fine-tuned for application to the eight-phase MR dataset. Conversely, a similar experiment was conducted where the model, initially trained on our eight-phase MR dataset, had its weights fine-tuned for the three-phase CT dataset. The experimental results outlined in Table \ref{tab:pretrain_compare}, demonstrate a notable performance improvement with the application of cross-modality transfer learning, underscoring the SDR-Former model's adaptability and scalability. Remarkably, utilizing pre-trained weights from the more complex, eight-phase MR dataset yielded more pronounced improvements than those from the simpler three-phase dataset, despite its smaller size. This suggests that the complex patterns gleaned from a dataset with a greater number of phases can enrich the model's knowledge base, thereby enhancing its performance even on datasets with fewer phases.
\begin{table}[h]
\centering
\caption{Efficacy of Transfer Learning.}
\label{tab:pretrain_compare}
\scalebox{1}{
    \begin{tabular}{lllll}
    \toprule
    SDR-Former & ACC   & AUC   & F1    & Kappa \\
    \midrule
    \textbf{CT (3-phase)} &       &       &       &  \\
    From Scratch & 0.8537  & 0.9168  & 0.8529  & 0.7058  \\
    MR pre-training & \textbf{0.8923 } & \textbf{0.9352 } & \textbf{0.8915 } & \textbf{0.7830 } \\
          &       &       &       &  \\
    \textbf{MR (8-phase)} &       &       &       &  \\
    From Scratch & 0.7885  & 0.9536  & 0.7910  & 0.7467  \\
    CT pre-training & \textbf{0.7981 } & 0.9350  & \textbf{0.8003 } & \textbf{0.7581 } \\
    \bottomrule
    \end{tabular}%
}
\end{table}
\subsection{Versatility Across Various Medical Datasets}
The DR-Former, the cornerstone of our framework, is designed as a universal vision backbone. In this section, we broaden the assessment of our model via extending its primary usage of multi-phase imaging. To be specific, we assessed the feature extraction capability of the DR-Former using MedMNIST v2, a recognized benchmark in medical image classification \cite{medmnistv2}, specifically focusing on subsets from MedMNIST3D, including OrganMNIST3D and NoduleMNIST3D. A distinctive feature of the DR-Former is its dual-resolution feature map generation. In our experiments, we down-scaled the higher-resolution maps to match the lower-resolution ones, then merged them. This combination was followed by processing through a GAP layer and a FC layer for classification. The comparative results presented in Table \ref{tab:medmnist_compare} consistently demonstrate the DR-Former's enhanced performance in comparison to other top-performing methods across multiple metrics. These findings highlight the DR-Former's robustness and establish it as an effective tool for medical image analysis, confirming its broad effectiveness beyond multi-phase imaging applications.
\begin{table}[h]
\centering
\caption{Comparative analysis of model performance on MedMNIST v2 dataset.}
\label{tab:medmnist_compare}
\scalebox{1}{
    \begin{tabular}{lllll}
    \toprule
    Method & ACC   & AUC   & F1    & Kappa \\
    \midrule
    \textbf{OrganMNIST3D} &       &       &       &  \\
    ResNet-50 & 0.9541  & 0.9988  & 0.9599  & 0.9489  \\
    DenseNet-121 & 0.9459  & 0.9978  & 0.9551  & 0.9397  \\
    MCSCNN & 0.9426  & 0.9983  & 0.9521  & 0.9361  \\
    BoTNet-50 & 0.9475  & 0.9987  & 0.9553  & 0.9415  \\
    UniFormer-S & 0.9279  & 0.9977  & 0.9391  & 0.9196  \\
    H2Former & 0.9410  & 0.9980  & 0.9495  & 0.9342  \\
    DR-Former & \textbf{0.9623 } & \textbf{0.9990 } & \textbf{0.9681 } & \textbf{0.9580 } \\
          &       &       &       &  \\
    \textbf{NoduleMNIST3D} &       &       &       &  \\
    ResNet-50 & 0.8871  & 0.8939  & 0.8267  & 0.6534  \\
    DenseNet-121 & 0.8806  & 0.9057  & 0.8168  & 0.6336  \\
    MCSCNN & 0.8806  & 0.9177  & 0.8229  & 0.6460  \\
    BoTNet-50 & 0.8742  & 0.8889  & 0.8022  & 0.5595  \\
    UniFormer-S & 0.8548  & 0.8521  & 0.7798  & 0.5602  \\
    H2Former & 0.8710  & 0.9190  & 0.8053  & 0.6107  \\
    DR-Former & \textbf{0.9032 } & \textbf{0.9216 } & \textbf{0.8430 } & \textbf{0.6865 } \\
    \bottomrule
    \end{tabular}%
}
\end{table}
\section{Conclusion}
\label{conclusion}
This study introduces the SDR-Former, an innovative architecture specifically developed for classifying liver lesions in 3D multi-phase CT and MR images. The model capitalizes on an SNN to concurrently process multiple phases, thereby enhancing the robustness of feature representation while maintaining scalability and computational efficiency. To enhance the feature representation power of the SNN, a hybrid DR-Former is incorporated. This DR-Former combines the capabilities of CNNs and Transformers to optimize both local details and global context understanding. In addition, a BCIM is incorporated to enable interaction between features from the CNN and Transformer branches. Furthermore, an APSM is utilized to strategically highlight key multi-phase features for accurate classification. Extensive testing on two clinical 3D multi-phase CT and MR datasets demonstrates our model's flexibility and effectiveness under various phase conditions. To contribute to research in multi-phase liver lesion analysis, we plan to release our MR dataset publicly.
\par
Despite its strengths, the SDR-Former does have limitations, particularly its dependence on an SNN for multi-phase input processing. While the model exhibits remarkable scalability, its effectiveness could be less pronounced in scenarios with fewer phases, such as dual-phase cases, where feature-level phase fusion might be more effective. However, the model’s ability to adapt to a range of phase conditions is a notable advantage, typically unachievable by feature-level fusion methods due to their high computational demands. Future developments will focus on refining the feature extraction process by integrating dynamic convolutions into the CNN branch \cite{chen2020dynamic}. This enhancement aims to produce phase-specific features within a weight-sharing framework, potentially leading to unique feature extractors for each imaging phase.

\section*{Ethical Approval}
The ethical approval for the utilization of CT dataset in this study has been granted by the Institutional Review Board (IRB) at Sir Run Run Shaw Hospital, Zhejiang University School of Medicine, Hangzhou, Zhejiang, China (IRB Approval No: 20200903-43). Similarly, the ethical approval for the utilization of MR dataset has obtained from the IRB at Ningbo Medical Center Lihuili Hospital, Ningbo, Zhejiang, China (IRB Approval No: KY2022SL446-01).

\section*{Data Availability}
The MR dataset is available at: \textcolor{blue}{https://bit.ly/3IyYlgN}.

\section*{Acknowledgement}
This work was supported in part by the National Key Research and Development Program of China (No. 2021ZD0113302).

\bibliographystyle{unsrt}

\bibliography{refs}

\end{document}